\documentclass[a4paper,11pt]{article}

\usepackage[latin1]{inputenc}
\usepackage[top=2.5cm, left=2.5cm, right=2cm, bottom=2.5cm]{geometry}
\usepackage{graphicx}
\usepackage[centertags]{amsmath}
\usepackage{amsfonts}
\usepackage{amssymb}
\usepackage{amsthm}
\usepackage{newlfont}
\usepackage{xcolor}
\usepackage{soul}
\usepackage{lscape}

\begin{document}
\begin{center}

{\Large \bf{Spin-3 fields in Mielke-Baekler gravity} }

\vspace{1cm}
J. R. B. Peleteiro\footnote{jbpelleteiro@hotmail.com}, \; C. E. Valc\'arcel\footnote{valcarcel.flores@gmail.com}

\vspace{.5cm}

$^{a}$\emph{Instituto de F\'isica, Universidade Federal da Bahia, C\^ampus Universit\'ario de Ondina, 40210-340, Salvador-BA, Brazil}.

\end{center}

\vspace{.25cm}

\begin{abstract}

Mielke-Baekler gravity consists in the usual Einstein-Hilbert action with a cosmological term and rotational and translational Chern-Simons terms with arbitrary couplings. For a particular choice of these couplings, we can obtain Einstein-Hilbert action, its teleparallel equivalent, and the exotic Witten's gravity.

In this work, we use the Chern-Simons formalism to generalize the three dimensional Mielke-Baekler gravity theory in order to introduce spin-3 fields. We study its asymptotic symmetries, black hole solution, and also analyse its canonical structure at its singular point.
\vspace{.5cm}


\end{abstract}

\section{Introduction}

Outside local sources, three-dimensional Einstein-Hilbert $\left(\mathrm{EH}\right)$ gravity has constant curvature and no local degrees of freedom. However, as shown by Brown and Henneaux \cite{Brown-Henneaux}, the $\mathrm{AdS}_3$ asymptotic symmetries are two copies of the Virasoro algebra. This result is the precursor of the $\mathrm{AdS_3/CFT_2}$ correspondence and also showed that three-dimensional gravity is far from trivial.

The study of the black hole solutions and asymptotic symmetries of three-dimensional gravity has not been exclusive to $\mathrm{EH}$ gravity. It has been extended to other three-dimensional models such as Conformal Gravity \cite{conformal}, Topologically Massive Gravity \cite{TMG01}, \cite{TMG02}, New Massive Gravity and General Massive Gravity \cite{NMG01}, \cite{NMG02}. These models have some common features: They can be written as a Chern-Simons-like action \cite{CSL}, and they have a vanishing torsion. The Mielke-Baekler $\left(\mathrm{MB}\right)$ gravity model \cite{MB01}, \cite{MB02} consists in the usual $\mathrm{EH}$ action with a cosmological term plus a rotational and a translational Chern-Simons $\left(\mathrm{CS}\right)$ term. For a particular choice of parameters, we can recover $\mathrm{EH}$ gravity, Teleparallel gravity, and the parity-odd ``exotic" Witten's gravity \cite{Witten01}. Therefore, $\mathrm{MB}$ gravity is a good laboratory to test the role of curvature and torsion in the $\mathrm{AdS_3/CFT_2}$ correspondence. In references \cite{Blagojevic01}-\cite{NHTorsion} the relation of $\mathrm{MB}$ gravity with the $\mathrm{CS}$ action, black hole solutions, canonical structure and asymptotic symmetries has been explored and a supersymmetric extension of $\mathrm{MB}$ gravity was constructed in \cite{susyMB}.

On the other hand, there has been a lot of interest in the study of higher-spin gravity models due to, between others, their non-linear dynamics \cite{Vasiliev} and their relation with String Theory \cite{String}. In three dimensions, higher-spin $\mathrm{EH}$ gravity can be written as the difference of two $\mathrm{CS}$ actions under the $\mathrm{SL(N)}\times\mathrm{SL(N)}$ group \cite{Blencowe}-\cite{Campoleoni2}. This construction was generalized to spin-3 topologically massive gravity \cite{spin3a}, \cite{spin3b} and $\mathrm{CS}$-like gravity \cite{spin3cs}.

In this work, we build the spin-3 $\mathrm{MB}$ gravity. Different from other spin-3 gravity models, we do not need to impose any constraints on torsions and therefore we can build a teleparallel equivalent spin-3 $\mathrm{EH}$ action. We also use its $\mathrm{CS}$ formulation to obtain black-hole solutions and asymptotic symmetries, which consists of two copies of the $W_3$ algebra with different $\mathrm{CS}$ central charges. The pure $\mathrm{MB}$ gravity theory has a singular point and we are going to show that this point is preserved in the presence of the spin-3 fields and analyse its canonical structure.

This paper is organized as follows: In section \ref{sec2}, we give a brief review of $\mathrm{MB}$ gravity and its $\mathrm{CS}$ formulation. In section \ref{sec3}, we use the $\mathrm{CS}$ formalism in order to introduce spin-3 fields in the $\mathrm{MB}$ gravity, analyse its equations of motion, show its black-hole solution, and also the asymptotic symmetries. In section \ref{sec4}, we present the singular point of the spin-3 $\mathrm{MB}$ theory and analyse its canonical structure using Dirac's algorithm. Finally, in section \ref{sec5}, we comment on our results and give future perspectives of this work.

\section{The $\mathrm{AdS}$ sector of Mielke-Baekler gravity theory} \label{sec2}
In the first-order formalism of three-dimensional gravity, the independent fields are the dreibein $e^a=e^a_\mu\mathrm{d}x^\mu$ and the dual spin connection $\omega^a=\omega^a_\mu\mathrm{d}x^\mu$. The curvature $R_a$ and torsion $T_a$ are defined by
\begin{equation}\label{00}
R_{a}=\mathrm{d}\omega_{a}+\frac{1}{2}\epsilon_{a}^{\;bc}\omega_{b}\wedge\omega_{c},\;\;\;\;\;\; T_{a}=\mathrm{d}e_{a}+\epsilon_{a}^{\;bc}\omega_{b}\wedge e_{c}.
\end{equation}
We lower and rise indices with the Minkowsky metric $\eta_{ab}=\mathrm{diag}\left(-1,1,1\right)$. The Levi-Civita symbol is denoted by $\epsilon^{abc}$ and by convention $\epsilon^{012}=1$.

The action for the three-dimensional Mielke-Baekler $\left(\mathrm{MB}\right)$ gravity theory \cite{MB01} is given by
\begin{equation}\label{01}
\mathrm{I_{MB}}=  \mathrm{a} \mathrm I_{1}+\Lambda \mathrm{I}_{2}+\alpha_{3}\mathrm{I}_{3}+\alpha_{4}\mathrm{I}_{4}
\end{equation}
where $\left(\mathrm{a},\Lambda,\alpha_3,\alpha_4\right)$ are free parameters and
\begin{eqnarray}
\mathrm{I}_{1}	&=&	2 \int\; e^{a}\wedge R_{a}\label{01a} \\
\mathrm{I}_{2}	&=&	-\frac{1}{3} \int\; \epsilon_{abc}e^{a}\;\wedge e^{b}\wedge e^{c}\label{01b} \\
\mathrm{I}_{3}	&=&	\int\; \left( \omega^{a}\wedge\mathrm{d}\omega_{a}+\frac{1}{3}\epsilon_{abc}\;\omega^{a}\wedge\omega^{b}\wedge\omega^{c}\right)\label{01c} \\
\mathrm{I}_{4}	&=&	\int\; e^{a}\wedge T_{a}\label{01d}.
\end{eqnarray}
The term $\mathrm{I}_{1}$ represents the Einstein-Hilbert $\left(\mathrm{EH}\right)$ action, $\mathrm{I}_{2}$ is a cosmological term, the third term $\mathrm{I}_{3}$ is a $\mathrm{CS}$ action for the spin-connection and $\mathrm{I}_{4}$ is a translational $\mathrm{CS}$ term.

The equations of motion of the $\mathrm{MB}$ action are:
\begin{eqnarray}
\mathrm{a} R_{a}+\alpha_{4}T_{a}-\frac{1}{2}\Lambda\epsilon_{abc}e^{b}\wedge e^{c}&=& 0\label{02}\\
\alpha_{3} R_{a}+\mathrm{a} T_{a}+\frac{1}{2}\alpha_{4}\epsilon_{abc}e^{b}\wedge e^{c} &=& 0.\label{03}
\end{eqnarray}
For an specific choice of parameters, the above equations degenerate to a single one \cite{MB02}. This is a singular point of the $\mathrm{MB}$ theory, and it has some similarities with the chiral point of Topological Massive Gravity \cite{Santamaria}.

For the non-degenerated sector, $\alpha_3\alpha_4-\mathrm{a}^2\neq0$, we can determine the curvature and torsion forms. The vacuum solution is characterized by a constant curvature
\begin{equation}\label{04}
R_{a}-\frac{\mathrm{q}}{2}\epsilon_{abc}e^{b}\wedge e^{c}=0,\;\;\;\;\;\;\;\;\; \mathrm{q}=-\frac{\mathrm{a}\Lambda+\left(\alpha_{4}\right)^{2}}{\alpha_{3}\alpha_{4}-\mathrm{a}^{2}}\end{equation}
and a constant torsion
\begin{equation}\label{05}
T_{a}-\frac{\mathrm{p}}{2}\epsilon_{abc}e^{b}\wedge e^{c}=0,\;\;\;\;\;\;\;\;\; \mathrm{p}=\frac{\alpha_{3}\Lambda+\alpha_{4}\mathrm{a}}{\alpha_{3}\alpha_{4}-\mathrm{a}^{2}}.
\end{equation}
We can identify three particular cases of the $\mathrm{MB}$ theory: First, the $\mathrm{EH}$ gravity with cosmological term. Here we set $\alpha_3=\alpha_4=0$, such that the torsion vanishes
\begin{equation}
R_{a}-\frac{\Lambda}{2\mathrm{a}}\epsilon_{abc}e^{b}\wedge e^{c}=0,\;\;\;\;\;\;\;\; T_{a}=0.\label{06}
\end{equation}

In addition, we fixed $\mathrm{a}=1/16\pi G$. Then, the curvature will be positive, negative or zero, depending on the value of $\Lambda$.

A second case corresponds to $\mathrm{a}\Lambda+\left(\alpha_{4}\right)^{2}=0$, where we have a vanishing curvature $\left(\mathrm{q}=0\right)$ and constant torsion
\begin{equation}
T_{a}+\frac{\alpha_4}{2\mathrm{a}}\epsilon_{abc}e^{b}\wedge e^{c} = 0,\;\;\;\;\;\;\;\; R_{a}=0.\label{07}
\end{equation}
This case corresponds to the three-dimensional Teleparallel theory. Note that the parameter $\alpha_3$ does not appears either in the field equations or in the condition for zero curvature, then, it can be set to zero without lost of generality.

The third case is the Witten's ``exotic" gravity \cite{Witten01}, where $\mathrm{a}=\Lambda=0$. This model have a vanishing torsion, a curvature that is proportional to $\mathrm{q}=-\alpha_4/\alpha_3$, and reversed roles for mass and angular momentum \cite{Townsend}.

It is convenient to write the dual spin connection in terms of the dual Levi-civita connection $\tilde\omega^a$ and the contorsion $\tau^a$: $\omega^a=\tilde\omega^a+\tau^a$. For $\mathrm{MB}$ gravity we have $\tau^{a}=\frac{\mathrm{p}}{2}e^{a}$, this can be directly computed from the field equation \eqref{05}. The curvature and torsion forms corresponding to the dual Levi-Civita connection are
\begin{equation}
\tilde{R}_{a}=\frac{1}{2}\Lambda_{\mathrm{eff}}\;\epsilon_{abc}e^{b}\wedge e^{c},\;\;\;\;\;\;\;\tilde{T}_i = 0,\;\;\;\;\;\Lambda_{\mathrm{eff}}\equiv \mathrm{q}-\frac{\mathrm{p}^{2}}{4}.
\end{equation}
In the above equation, $\Lambda_{\mathrm{eff}}$ acts as an effective cosmological constant. We can further restrict ourselves to the $\mathrm{AdS}$ sector, where $\Lambda_{\mathrm{eff}}$ is negative
\begin{equation}
\Lambda_{\mathrm{eff}}=  \mathrm{q}-\frac{\mathrm{p}^{2}}{4} = -\frac{1}{l^2} < 0.
\end{equation}
In the following table we show a choice of parameters $\left(\mathrm{a},\Lambda,\alpha_3,\alpha_4\right)$ for three particular cases which belong to the $\mathrm{AdS}$ sector of the $\mathrm{MB}$ theory:
\begin{eqnarray*}
 \renewcommand{\arraystretch}{1.7}
\begin{array}{| c | c | c | c | c | c | c |}
\hline  & \mathrm{a} & \Lambda & \alpha_{3} & \alpha_{4} & \mathrm{q} & \mathrm{p}\\
\hline \mathrm{Einstein-Hilbert} & \frac{1}{16\pi G} & -\frac{1}{16\pi G l^{2}} & 0 & 0 & -\frac{1}{l^{2}} & 0\\
\hline \mathrm{Teleparallel} & \frac{1}{16\pi G} & -\frac{1}{4\pi G l^2} & 0 & -\frac{1}{8\pi G l}  & 0 & \frac{2}{l} \\
\hline \mathrm{Exotic} & 0 & 0 & -\frac{l}{16\pi G} & -\frac{1}{16\pi Gl} & -\frac{1}{l^{2}} & 0
\\\hline \end{array}
\end{eqnarray*}
The values for the exotic theory are in agreement with \cite{Exotic}.

\subsection{Chern-Simons formulation}

In \cite{Blagojevic01} it was shown that the $\mathrm{AdS}$ sector of $\mathrm{MB}$ theory can be written as the difference of two $\mathrm{CS}$ actions with different $\mathrm{CS}$ levels. To achieve this, we need to define two Lie algebra valued gauge connections $A^\pm$ as the linear combination of the dual spin connection and the dreibein:
\begin{equation}\label{CS01}
A^{\pm} = A^a_\pm J^\pm_a=\left[\omega^a+\left(-\frac{\mathrm{p}}{2}\pm\frac{1}{l}\right)e^a\right]J^\pm_a.
\end{equation}
where $J^\pm_a$ are $\mathfrak{so}(2,1)\sim\mathfrak{sl}(2)$ generators: $\left[J^\pm_a,J^\pm_b\right]=\epsilon_{ab}^{\;\;\;c} J^\pm_c$. Then, we have that the $\mathrm{AdS}$ sector of $\mathrm{MB}$ can be written as
\begin{equation}\label{CS02}
\mathrm{I_{MB}}=\mathrm{I^+_{CS}}\left[A^{+}\right]-\mathrm{I^-_{CS}}\left[A^{-}\right]
\end{equation}
where
\begin{equation}\label{CS03}
\mathrm{I^\pm_{CS}}\left[A^{\pm}\right]	=	 \frac{k_{\pm}}{4\pi}\int_{\mathcal{M}}\mathrm{tr}\left[A^{\pm}\wedge\mathrm{d}A^{\pm}+\frac{2}{3}A^{\pm}\wedge A^{\pm}\wedge A^{\pm}\right]+\mathrm{B}^\pm\left[A^\pm\right].
\end{equation}
In the above equation, $\mathrm{B}^\pm$ are boundary terms, the traces are $\mathrm{tr}\left(J^\pm_a J^\pm_b\right)=\eta_{ab}/2$, and $\mathrm{tr}\left(J^\pm_a J^\mp_b\right)=0$. The $\mathrm{CS}$ levels $k_{\pm}$ are given by:
\begin{equation}\label{CS04}
k_{\pm} = 4\pi \left[\left(\mathrm{a}+\alpha_{3}\frac{\mathrm{p}}{2}\right)l\pm\alpha_{3}\right].
\end{equation}
For $\mathrm{EH}$ and Teleparallel gravity, the $\mathrm{CS}$ levels become equal to $k_+=k_-= l/4G$. For ``exotic" gravity, the $\mathrm{CS}$ leves are $k_\pm = \mp l/4G$. Furthermore, the $\mathrm{CS}$ action \eqref{CS02} has the equation of motion
\begin{equation}\label{CS05}
F^\pm = \mathrm{d}A^\pm + A^\pm\wedge A^\pm=0
\end{equation}
and it is invariant under the gauge transformation
\begin{equation}\label{CS06}
\delta_{\lambda}A^{\pm}=\mathrm{d}\lambda^{\pm}+\left[A^{\pm},\lambda^{\pm}\right]. 
\end{equation}
The importance of the $\mathrm{CS}$ formulation lies in that we have a machinery to explore boundary conditions, black hole solutions and asymptotic symmetries \cite{Banados01}, \cite{Banados02}, \cite{Miskovic1}. In the next section we will make use the $\mathrm{CS}$ action to couple $\mathrm{MB}$ with spin-3 fields.

\section{Spin-3 generalization of Mielke-Baekler gravity}\label{sec3}

Three-dimensional $\mathrm{EH}$ gravity coupled with symmetric tensor fields of spin $\mathrm{N}$ can be written as the difference of two $\mathrm{CS}$ actions under the group $\mathrm{SL(N)}\times\mathrm{SL(N)}$ \cite{Campoleoni}. Analogously, we state that the spin $\mathrm{N}$ Mielke-Baekler theory is the difference of two $\mathrm{CS}$ action under the group $\mathrm{SL}\left(\mathrm{N}\right)\otimes\mathrm{SL}\left(\mathrm{N}\right)$ with $\mathrm{CS}$ levels given by $\eqref{CS04}$ \footnote{In \cite{Campoleoni2} was also proposed a higher-spin generalization of $\mathrm{MB}$ gravity. However, we will not restrict to the $\mathrm{AdS}$ sector, as we will see in the next section}.

In the following we consider $\mathrm{N}=3$ in order to  have an explicit expression for the higher-spin Mielke-Baekler action. In this case, the gauge connection is given by
\begin{equation}\label{HS01}
A^{\pm}=\left[\omega^{a}+\left(-\frac{\mathrm{p}}{2}\pm\frac{1}{l}\right)e^{a}\right]J_{a}^{\pm}+\left[\omega^{ab}+\left(-\frac{\mathrm{p}}{2}\pm\frac{1}{l}\right)e^{ab}\right]T_{ab}^{\pm}
\end{equation}
where $e^{ab}$ is a vielbein-like field and $\omega^{ab}$ an auxiliary field. These new quantities are traceless and symmetric in the flat indices. The generators of the $\mathfrak{sl}(3)$ algebra, denoted by $J_{a}^{\pm},T_{ab}^{\pm}$, satisfy
\begin{eqnarray}
 \left[J_{a}^{\pm},J_{b}^{\pm}\right]	&=&	\epsilon_{ab}^{\;\;\; c}J^\pm_{c}\label{HS01a}\\
\left[J_{a}^{\pm},T_{bc}^{\pm}\right]	&=&	\epsilon^d_{\;\; a(b}T^\pm_{c)d}\label{HS01b}\\
\left[T_{ab}^{\pm},T_{cd}^{\pm}\right]	&=&	\sigma\left[\eta_{a(c}\epsilon_{d)b}^{\;\;\;\;\; e}+\eta_{b(c}\epsilon_{d)a}^{\;\;\;\;\; e}\right]J^\pm_{e}\label{HS01c}
\end{eqnarray}
where a pair of parentheses on the indices denotes a complete symmetrization without normalization factor, and $\sigma$ is a negative parameter (Positive values of $\sigma$ are related to the $\mathfrak{su}(1,2)$ algebra). Furthermore, we have that
\begin{equation}
\mathrm{tr}\left(J_{a}^{\pm}J_{b}^{\pm}\right)=\frac{1}{2}\eta_{ab},\;\;\mathrm{tr}\left(J_{a}^{\pm}T_{bc}^{\pm}\right)=0,\;\;\mathrm{tr}\left(T_{ab}^{\pm}T_{cd}^{\pm}\right)=-\frac{\sigma}{2}\left(\eta_{a(c}\eta_{d)b}-\frac{2}{3}\eta_{ab}\eta_{cd}\right).
\end{equation}

Replacing \eqref{HS01} in \eqref{CS02}, we obtain the spin-3 Mielke-Baekler action
\begin{equation}\label{HS02}
\mathrm{I^{Spin-3}_{MB}}= \mathrm{a} \mathrm{I}^{(3)}_1+ \Lambda \mathrm{I}^{(3)}_2 + \alpha_3 \mathrm{I}^{(3)}_3+\alpha_4 \mathrm{I}^{(3)}_4
\end{equation}
where
\begin{eqnarray}
\mathrm{I}^{(3)}_1 &=& 2 \int\left( e^{a}\wedge R^{(3)}_{a}-2\sigma e^{ab}\wedge R^{(3)}_{ab} \right) \label{HS02a}\\
\mathrm{I}^{(3)}_2 &=& -\frac{1}{3}\int \left(\epsilon_{abc}e^{a}\wedge e^{b}\wedge e^{c}-12\sigma\epsilon_{abc}e^{a}\wedge e_{\; d}^{b}\wedge e^{dc}\right)\label{HS02b}\\
\mathrm{I}^{(3)}_3 &=& \int \left(\omega^{a}\wedge\mathrm{d}\omega_{a}+\frac{1}{3}\epsilon_{abc}\omega^{a}\wedge\omega^{b}\wedge\omega^{c}-2\sigma\omega^{ab}\wedge\mathrm{d}\omega_{ab}-4\sigma\epsilon_{abc}\omega^{a}\wedge\omega_{\; d}^{b}\wedge\omega^{dc}\right) \label{HS02c}\\
\mathrm{I}^{(3)}_4 &=& \int \left( e^{a}\wedge T^{(3)}_{a}-2\sigma e^{ab}\wedge T^{(3)}_{ab} \right).\label{HS02d}
\end{eqnarray}
These terms are the spin-3 generalization of \eqref{01a}-\eqref{01d}, respectively. We still have four couplings: $(\mathrm{a},\Lambda,\alpha_3,\alpha_4)$, reminiscent of the supersymmetric generalization \cite{susyMB}. However, they are constrained to the condition $\mathrm{a}^2-\alpha_3\alpha_4\neq0$ since the connection \eqref{HS01} depends on $l,\mathrm{p},\mathrm{q}$. The curvatures are defined by
\begin{eqnarray}
R^{(3)}_{a}	&\equiv&	 \mathrm{d}\omega_{a}+\frac{1}{2}\epsilon_{abc}\omega^{b}\wedge\omega^{c}-2\sigma\epsilon_{abc}\omega_{\; d}^{b}\wedge\omega^{dc}\label{hs03a}\\
R^{(3)}_{ab}	&\equiv&	\mathrm{d}\omega_{ab}+\epsilon_{cd(a}\omega^{c}\wedge\omega_{b)}^{\;d}\label{hs03b}
\end{eqnarray}
and the torsions are given by
\begin{eqnarray}
T^{(3)}_{a}	&\equiv&	\mathrm{d}e_{a}+\epsilon_{abc}\omega^{b}\wedge e^{c}-4\sigma\epsilon_{abc}e^{bd}\wedge\omega_{d}^{\; c}\label{hs04a}\\
T^{(3)}_{ab}	&\equiv&	\mathrm{d}e_{ab}+\epsilon_{cd(a}\omega^{c}\wedge e_{b)}^{\;\;d}+\epsilon_{cd(a}e^{c}\wedge\omega_{b)}^{\;d}.\label{hs04b}
\end{eqnarray}
Note that in the absence of spin-3 fields, the quantities $R^{(3)}_a$ and $T^{(3)}_a$ coincide with the usual curvature and torsion \eqref{00} and $R^{(3)}_{ab}=T^{(3)}_{ab}=0$. Also notice that, for $\alpha_3=\alpha_4=0$ and $\mathrm{a}=-l\Lambda=1/16\pi G$, the action \eqref{HS02} reduces to the usual spin-3 $\mathrm{EH}$ action with cosmological constant.

The variation of the action with respect to $e_a$ and $\omega_a$ fields gives the following equations of motion
\begin{eqnarray}
\mathrm{a}R^{(3)}_{a}+\alpha_{4}T^{(3)}_{a}-\Lambda\left[ \frac{1}{2}\epsilon_{abc}e^{b}\wedge e^{c}-2\sigma\epsilon_{abc}e^{bd}\wedge e_d^{\;c} \right] &=& 0 \label{hs05a}\\
\alpha_{3}R^{(3)}_{a}+\mathrm{a}T^{(3)}_{a}+\alpha_{4}\left[\frac{1}{2}\epsilon_{abc}e^{b}\wedge e^{c}-2\sigma\epsilon_{abc}e^{bd}\wedge e_{d}^{\; c}\right]	&=&	0 \label{hs05b}
\end{eqnarray}
which are generalizations of \eqref{02}, \eqref{03}. The variations with respect to $e_{ab},\omega_{ab}$ give
\begin{eqnarray}
\mathrm{a}R^{(3)}_{ab}+\alpha_{4}T^{(3)}_{ab}-\Lambda\epsilon_{cd(a}e^{c}\wedge e_{b)}^{\; d} &=& 0 \label{hs05c}\\
\alpha_{3}R^{(3)}_{ab}+\mathrm{a}T^{(3)}_{ab}+\alpha_{4}\epsilon_{cd(a}e^{c}\wedge e_{b)}^{\; d} &=& 0. \label{hs05d}
\end{eqnarray}
Equations \eqref{hs05a}-\eqref{hs05d} are equivalent to zero curvature equation \eqref{CS05} in the $\mathrm{CS}$ formalism.

Now, since we build the spin-3 $\mathrm{MB}$ theory from the $\mathrm{CS}$ formulation, the condition $\mathrm{a}^2-\alpha_3\alpha_4\neq0$ holds. Then, the two sets of equations above can be solved for the curvatures and torsions:
\begin{eqnarray}
\mathcal{R}^{(3)}_a &\equiv& R^{(3)}_{a}-\frac{1}{2}\mathrm{q}\epsilon_{abc}e^{b}\wedge e^{c}+2\sigma\mathrm{q}\epsilon_{abc}e_{\; m}^{b}\wedge e^{mc}=0\label{hs06a}\\
\mathcal{T}^{(3)}_a &\equiv& T^{(3)}_{a}-\frac{1}{2}\mathrm{p}\epsilon_{abc}e^{b}\wedge e^{c}+2\sigma\mathrm{p}\epsilon_{abc}e_{\; m}^{b}\wedge e^{mc}=0\label{hs06b}\\
\mathcal{R}^{(3)}_{ab} &\equiv& R^{(3)}_{ab}-\mathrm{q}\epsilon_{cd(a}e^{c}\wedge e_{b)}^{\; d}=0\label{hs06c}\\
\mathcal{T}^{(3)}_{ab} &\equiv& T^{(3)}_{ab}- \mathrm{p}\epsilon_{cd(a}e^{c}\wedge e_{b)}^{\; d}=0. \label{hs06d}
\end{eqnarray}
Note that for spin-3 $\mathrm{EH}$ or exotic gravity, where $\mathrm{p}=0$, we have the zero torsion constraints: $T^{(3)}_{a}=0=T^{(3)}_{ab}$. However, the torsion defined in \eqref{00} is not zero, which resemble the supergravity generalization of $\mathrm{MB}$ gravity \cite{susyMB}. On the other hand, for the spin-3 teleparallel gravity $\mathrm{q}=0$, we have that the curvatures are zero $R^{(3)}_{a}=0=R^{(3)}_{ab}$. In the next section, we will show that the spin-3 teleparallel gravity has the same central charge that the spin-3 $\mathrm{EH}$ action. In this sense, they are equivalent.

Now, for the pure $\mathrm{MB}$ gravity theory, the $\mathrm{AdS}$ sector can be explicitly realized when we write the spin-connection as the sum of the Levi-Civita connection and a contorsion. Analogously, we can perform a map that leads us from the family of the four parametric model to the $\mathrm{AdS}$ sector with and effective cosmological constant. This transformation is given by
\begin{equation}
\omega_{a}=\tilde{\omega}_{a}+\frac{\mathrm{p}}{2}e_{a},\;\;\;\;\omega_{ab}=\tilde{\omega}_{ab}+\frac{\mathrm{p}}{2}e_{ab}
\end{equation}
where $\tilde{\omega}_{a}$ and $\tilde{\omega}_{ab}$ satisfy $T_{a}^{\left(3\right)}\left(\tilde{\omega}\right)=0=T_{ab}^{\left(3\right)}\left(\tilde{\omega}\right)$. The remaining equations of motion become
\begin{eqnarray}
R^{(3)}_{a}\left(\tilde{\omega}\right) - \frac{1}{l^{2}}\left(\frac{1}{2}\epsilon_{abc}e^{b}\wedge e^{c}-2\sigma\epsilon_{abc}e_{\; d}^{b}\wedge e^{dc}\right) &=& 0\\
R^{(3)}_{ab}\left(\tilde{\omega}\right) + \frac{1}{l^{2}}\epsilon_{cd(a}e^{c}\wedge e_{b)}^{\; d} &=& 0
\end{eqnarray}
which are exactly the equations of motion for spin-3 $\mathrm{EH}$ action with and effective cosmological constant $-l^{-2}$.

\subsection{Asymptotic Symmetries and Spin-3 $\mathrm{MB}$ Black Hole}

The asymptotic symmetries and black hole solution of pure $\mathrm{MB}$ gravity has been explored using the $\mathrm{CS}$ formulation in  \cite{Blagojevic02}, \cite{Blagojevic03}. The same machinery can be used for the spin-3 $\mathrm{MB}$ theory. We will closely follow the spin-3 $\mathrm{EH}$ case found in \cite{Generalized}. First, let us consider the manifold $\mathcal M$ with the topology of a cylinder with radial, angular and temporal coordinates denoted by $\left(\rho, \varphi, t\right)$, respectively. The equation of motion \eqref{CS05} allows the following decomposition of the gauge connection:
\begin{equation}\label{ASA02}
A^\pm = b^{-1}_\pm\left(\rho\right)\left[\mathrm{d}+\mathrm{a}^\pm_\varphi\left(t,\varphi\right)\mathrm{d}\varphi + \mathrm{a}^\pm_t \left(t,\varphi\right)\mathrm{d}t \right]b_{\pm}\left(\rho\right)
\end{equation}
where $b_\pm$ is a group element and $\mathrm{a}^\pm_{t,\varphi}$ are auxiliary connections. Before specify the explicit form of these connections, it is convenient to transform the generators from $\left(J_a,J_{ab}\right)$ to the $\left(L^\pm_i,W^\pm_m\right)$ basis, where
\begin{eqnarray}
\left[L^\pm_i,L^\pm_j\right] &=& \left(i-j\right)L^\pm_{i+j}\\
\left[L^\pm_i,W^\pm_m\right] &=& \left(2i-m\right)W^\pm_{i+m}\\
\left[W^\pm_m,W^\pm_n\right] &=& \frac{\sigma}{3}\left(m-n\right)\left(2m^2+2n^2-mn-8\right)L^\pm_{m+n}
\end{eqnarray}
and $i,j=\pm1,0$, and $m,n=\pm2,\pm1,0$.
Now, we impose the following boundary conditions on the angular component
\begin{eqnarray}\label{BH03}
a^\pm_\varphi \left(t,\varphi\right) &=&  L_{\pm1}-\frac{2\pi}{k_{\pm}}\mathcal{L}_{\pm}L_{\mp1}+\frac{\pi}{2k_{\pm}\sigma}\mathcal{W}_{\pm}W_{\mp2}
\end{eqnarray}
and for the temporal component, we write $a_{t}^{\pm}=\Lambda^\pm\left(\mu^\pm,\nu^\pm\right)$, where:
\begin{eqnarray}\label{BH04}
\Lambda^\pm \left(\mu^\pm,\nu^\pm\right)	&\equiv&	 \pm\mu_{\pm}L_{\pm1}-\mu'_{\pm}L_{0}\pm\frac{1}{2}\left(\mu''_{\pm}-\frac{4\pi}{k_{\pm}}\mathcal{L}_{\pm}\mu_{\pm}+\frac{8\pi}{k_{\pm}}\mathcal{W}_{\pm}\nu_{\pm}\right)L_{\mp1}\pm\nu_{\pm}W_{\pm2}-\nu'_{\pm}W_{\pm1}\nonumber\\
&&\pm\left[\frac{1}{24}\nu_{\pm}^{\left(4\right)}-\frac{4\pi}{3k_{\pm}}\mathcal{L}_{\pm}\nu''_{\pm}-\frac{7\pi}{6k_{\pm}}\mathcal{L}'_{\pm}\nu'_{\pm}-\frac{\pi}{3k_{\pm}}\mathcal{L}''_{\pm}\nu_{\pm}+\frac{4\pi^{2}}{k_{\pm}^{2}}\mathcal{L}^2_{\pm}\nu_{\pm}+\frac{\pi}{2k_{\pm}\sigma}\mathcal{W}_{\pm}\mu_{\pm}\right]W_{\mp2}\nonumber\\
&& \pm\frac{1}{2}\left(\nu''_{\pm}-\frac{8\pi}{k_{\pm}}\mathcal{L}_{\pm}\nu_{\pm}\right)W_{0}-\frac{1}{6}\left(\nu'''_{\pm}-\frac{20\pi}{k_{\pm}}\mathcal{L}_{\pm}\nu'_{\pm}-\frac{8\pi}{k_{\pm}}\mathcal{L}'_{\pm}\nu_{\pm}\right)W_{\mp1}.
\end{eqnarray}
The function $\Lambda$ depends of the charges $\mathcal L_\pm, \mathcal W_\pm$ (arbitrary functions of $t,\varphi$) and the so called chemical potentials $\mu^\pm,\nu^\pm$. In the above equation the prime denotes angular derivatives. The boundary conditions \eqref{BH03}, \eqref{BH04} are in agreement with the equation of motion \eqref{CS05}.

The gauge parameter in \eqref{CS06} can be decomposed as $\lambda^\pm = b^{-1}_\pm(\rho) \eta^\pm(t,\varphi) b_\pm(\rho) $. The asymptotic form of $\eta$ that preserves the boundary condition \eqref{BH03} is $\eta^\pm = \Lambda^\pm(\epsilon^\pm,\xi^\pm)$, provided that the charges transform as
\begin{eqnarray}
\delta_{\epsilon,\xi}\mathcal{L}_{\pm} &=&	 \mp\frac{k_{\pm}}{4\pi}\epsilon'''_{\pm}\pm2\mathcal{L}_{\pm}\epsilon'_{\pm}\pm\mathcal{L}'_{\pm}\epsilon_{\pm}\mp3\mathcal{W}_{\pm}\xi'_{\pm}\mp2\mathcal{W}'_{\pm}\xi_{\pm}\label{BH05a}\\
\delta_{\epsilon,\xi}\mathcal{W}_{\pm}	&=&	 \pm\frac{k_{\pm}\sigma}{12\pi}\nu_{\pm}^{\left(5\right)}\mp\frac{10}{3}\sigma\mathcal{L}_{\pm}\nu'''_{\pm}\mp5\sigma\mathcal{L}'_{\pm}\nu''_{\pm}\mp3\sigma\left(\mathcal{L}''_{\pm}-\frac{64\pi}{9k_{\pm}}\mathcal{L}_{\pm}^{2}\right)\xi'_{\pm}\nonumber\\
&&\mp\frac{2\sigma}{3}\left(\mathcal{L}'''_{\pm}-\frac{16\pi}{k_{\pm}}(\mathcal{L}_{\pm}^{2})'\right)\xi_{\pm}\pm\mathcal{W}'_{\pm}\epsilon_{\pm}\pm3\mathcal{W}_{\pm}\epsilon'_{\pm}\label{BH05b}
\end{eqnarray}
where $\epsilon^\pm,\xi^\pm$ are arbitrary functions. We can decompose the charges $\mathcal L,\mathcal W$ in Fourier modes, and show that equations \eqref{BH05a}, \eqref{BH05b} represent two copies of the $\mathrm{W}_3$ algebra with two different central charges
\begin{equation}
    c_\pm = 6 k_\pm = 24\pi \left[\left(\mathrm{a}+\alpha_{3}\frac{\mathrm{p}}{2}\right)l\pm\alpha_{3}\right].\label{BH06}
\end{equation}
Since the values of $c_\pm$ for spin-3 $\mathrm{EH}$ and Teleparallel gravity coincide, as in the pure $\mathrm{MB}$ case, we say that the teleparallel model is equivalent to the $\mathrm{EH}$ action in the sense that the curvatures $R^{(3)}_a,R^{(3)}_{ab}$ are zero.

Now, the black-hole solution are obtained when the charges and chemical potentials are considered constants. Under these condition the temporal auxiliary component reduces to
\begin{eqnarray}
 a_{t}^{\pm}&=&	 \pm\mu_{\pm}\left(L_{\pm1}-\frac{2\pi}{k_{\pm}}\mathcal{L}_{\pm}L_{\mp1}+\frac{\pi}{2k_{\pm}\sigma}\mathcal{W}_{\pm}W_{\mp2}\right)\nonumber\\
&&\pm\nu_{\pm}\left(W_{\pm2}+\frac{4\pi}{k_{\pm}}\mathcal{W}_{\pm}L_{\mp1}-\frac{4\pi}{k_{\pm}}\mathcal{L}_{\pm}W_{0}+\frac{4\pi^{2}}{k_{\pm}^{2}}\mathcal{L}^2_{\pm}W_{\mp2}\right).\label{BH07}
\end{eqnarray}
Note that for $\nu_\pm=0$, we have that the temporal and angular components are proportional, this is a common choice in holography since it allows to integrate the boundary term on the action, and also allows to redefine the boundary condition on the null-directions \cite{Campoleoni}, \cite{BH301}, \cite{BH302}. 

In order to compute thermodynamic properties of the black hole we need to perform a Wick rotation $t\rightarrow it_E$. Then, the topology of the Euclidean black hole is equivalent to a torus and the light-cone coordinates $x^\pm = i t_E \pm\varphi$ can be identified with the independent complex variables $x^+\rightarrow z,\;x^{-}\rightarrow-\bar{z}$ where $(z,\bar z) \simeq (z+2\pi\tau,\bar z + 2\pi \bar \tau)$ and $\tau$ is the unimodular parameter of the torus. We are interested in regular black hole solutions, this is achieved when the holonomies
\begin{equation}
\mathrm{Hol}_{\tau,\bar{\tau}}\left(A^{+}\right)=b^{-1}\exp2\pi\left[\tau a_{z}^{+}+\bar{\tau}a_{\bar{z}}^{+}\right]b,\;\;\mathrm{Hol}_{\tau,\bar{\tau}}\left(A^{-}\right)=b\exp2\pi\left[\tau a_{z}^{-}+\bar{\tau}a_{\bar{z}}^{-}\right]b^{-1}.\label{BH09}
\end{equation}
are trivial. This implies that (see \cite{BH301}, \cite{BH302}):
\begin{equation}
\mathrm{Eigen}\left[2\pi\left(\tau a_{z}^{+}+\bar{\tau}a_{\bar{z}}^{+}\right)\right] = \mathrm{Eigen}\left[2\pi\left(\tau a_{z}^{-}+\bar{\tau}a_{\bar{z}}^{-}\right)\right]=\mathrm{Eigen}[2\pi i L_0].\label{BH10}
\end{equation}
In the next section we will compute these holonomies and obtain the entropy of the $\mathrm{MB}$ black hole solution.

\subsection{Entropy of the Mielke-Baekler black hole:}

In \cite{BanadosHS} was showed that the $\mathrm{SL(N)}$ $\mathrm{CS}$ action is finite. Then, we can compute the free energy as the on-shell action times temperature and perform a Legendre transformation to obtain the entropy. In \cite{Thermo} it was given a simple expression for the entropy in terms of the complex variables. For two different $\mathrm{CS}$ levels the entropy is given by
\begin{equation}
S=-2\pi i\;\mathrm{tr}\left[\tilde{k}_{+}\left(a_{z}^{+}+a_{\bar{z}}^{+}\right)\left(\tau a_{z}^{+}+\bar{\tau}a_{\bar{z}}^{+}\right)-\tilde{k}_{-}\left(a_{z}^{-}+a_{\bar{z}}^{-}\right)\left(\tau a_{z}^{-}+\bar{\tau}a_{\bar{z}}^{-}\right)\right]\label{BH11}
\end{equation}
where $\tilde{k}_\pm = k_\pm/[2\;\mathrm{tr_f}\left(L_{0}L_{0}\right)]$ and $\mathrm{tr_f}$ is the trace in the fundamental representation. The overall factor is chosen such that the entropy will match the normalization of $\mathrm{MB}$ gravity.

Before going to the $\mathrm{MB}$ gravity coupled with spin-3 fields, let us first consider the spin-2 case which corresponds to switch off the spin-3 charges $\mathcal W_\pm$ and set $\mu_\pm=1,\;\nu_\pm=0$. This choice of chemical potentials represents the $\mathrm{BTZ}$ black hole with torsion. From the matrix representation for the $\mathfrak{sl}(2)$ algebra (see Appendix) we have that $\tilde k_\pm = k_\pm$ and the entropy \eqref{BH11} becomes
\begin{equation}
S_{\mathrm{spin-2}}=4\pi\beta\left[\left(1+\Omega\right)\mathcal{L}_{+}+\left(1-\Omega\right)\mathcal{L}_{-}\right]\label{BH12}
\end{equation}
where $\beta=-i\pi\left(\tau-\bar{\tau}\right)$ and $\Omega=-\frac{i\pi}{\beta}\left(\tau+\bar{\tau}\right)$. The holonomy conditions lead us to
\begin{equation}
\beta\left(1\pm\Omega\right)=\sqrt{ \frac{\pi k_{\pm}}{2\mathcal{L}_{\pm}}} \label{BH12b}
\end{equation}
which is equivalent to determine $\tau$ and $\bar \tau$. Replacing in \eqref{BH12} we obtain
\begin{equation}
S_{\mathrm{spin-2}}=2\pi\left[\sqrt{2\pi k_{+}\mathcal{L}_{+}}+\sqrt{2\pi k_{-}\mathcal{L}_{-}}\right]\label{BH13}
\end{equation}
This result was first derived in \cite{Blagojevic05}. In this procedure we considered that $k_\pm>0$. However, if one of these parameters are negative, for example $k_-$, we have to consider its absolute value and write a minus sign in front of the second term in \eqref{BH13}, as noticed in \cite{Exo} for exotic gravity.

It is customary to write the entropy \eqref{BH13} in terms of the two horizons $r_\pm$ of its $\mathrm{BTZ}$ black hole. We write $\mathcal{L}_{\pm}=\frac{k_{\pm}}{\pi}\left(\mathcal{M}\pm\frac{1}{l}\mathcal{J}\right)$ and $r \equiv	 l\left[e^{2\rho}+4\mathcal{M}+4\left(\mathcal{M}^{2}-\frac{1}{l^{2}}\mathcal{J}^{2}\right)e^{-2\rho}\right]^{1/2}$. Therefore, for $\mathrm{EH}$ and its teleparallel equivalent we obtain $S_{\mathrm{spin-2}}=\frac{\pi r_{+}}{2G}$ and for exotic gravity $S_{\mathrm{spin-2}}=\frac{\pi r_{-}}{2G}$. Its important to stress that the quantities $\mathcal M, \mathcal J$ coincide with the mass and angular momentum of the black hole for $\mathrm{EH}$ gravity but they exchange roles for exotic gravity.

Now let us work for the spin-3 case. In this case we have $\tilde k_\pm = k/4$ and, in analogy with the spin-2 case, let us consider $\mu_\pm=1$. Replacing the black hole solution in the entropy, we obtain
\begin{equation}
S_{\mathrm{spin-3}}=4\pi\beta\left[\left(1+\Omega\right)\mathcal{L}_{+}-\frac{3}{2}\mathcal{W}_{+}\nu_{+}+\left(1-\Omega\right)\mathcal{L}_{-}-\frac{3}{2}\mathcal{W}_{-}\nu_{-}\right]\label{BH14}
\end{equation}
The holonomy conditions \eqref{BH10} are:
\begin{eqnarray}
2^{11}\pi^{2}\nu_{\pm}^{3}\mathcal{L}_{\pm}^{3}+3^{3}k_{\pm}^{2}\left(1\pm\Omega\right)^{3}\mathcal{W}_{\pm}-2^{5}3^{3}k_{+}\pi\nu_{\pm}^{3}\mathcal{W}_{\pm}^{2}&&\nonumber\\
+2^{5}3^{3}k_{\pm}\pi\left(1\pm\Omega\right)\nu_{\pm}^{2}\mathcal{L}_{\pm}\mathcal{W}_{\pm}-2^{6}3^{2}k_{\pm}\pi\left(1\pm\Omega\right)^{2}\nu_{\pm}\mathcal{L}_{\pm}^{2} &=& 0\label{BH16}
\end{eqnarray}
and
\begin{equation}
\frac{2^{6}}{3k_{\pm}^{2}}\beta^{2}\nu_{\pm}^{2}\mathcal{L}_{\pm}^{2}+\frac{2}{k_{\pm}\pi}\beta^{2}\left(1\pm\Omega\right)^{2}\mathcal{L}_{\pm}-\frac{6}{k_{\pm}\pi}\beta^{2}\left(1\pm\Omega\right)\nu_{\pm}\mathcal{W}_{\pm}-1=0.\label{BH17}
\end{equation}
In the above expressions we set $\sigma=-1$ to avoid clutter. The system of equations \eqref{BH16}-\eqref{BH17} are similar to ones for spin-3 $\mathrm{EH}$ gravity \cite{Generalized} and spin-3 flat space \cite{Matulich} and can be parameterized by two angles $\Phi_\pm$. We write $\mathcal{W}_{\pm}=\frac{2^{3}}{3}\sqrt{\frac{2\pi\mathcal{L}_{\pm}^{3}}{3k_{\pm}}}\sin\Phi_{\pm}$. Then
\begin{equation}
\beta\nu_{\pm}=\frac{\sqrt{3}k_{\pm}}{2^{3}\mathcal{L}_{\pm}}\frac{\sin\left(\Phi_{\pm}/3\right)}{\cos\Phi_{\pm}},\;\;\;\beta\left(1\pm\Omega\right)=\sqrt{\frac{k_{\pm}\pi}{2\mathcal{L}_{\pm}}}\frac{\cos\left(2\Phi_{\pm}/3\right)}{\cos\Phi_{\pm}}.   \label{BH18}
\end{equation}
In the absence of spin-3 charges, the solution \eqref{BH18} coincides with \eqref{BH12b}. Finally, the entropy becomes
\begin{equation}
S=2\pi\sqrt{2\pi k_{+}\mathcal{L}_{+}}\cos\left(\Phi_{+}/3\right)+2\pi\sqrt{2\pi k_{-}\mathcal{L}_{-}}\cos\left(\Phi_{-}/3\right)
\end{equation}
which is the spin-3 generalization of entropy for $\mathrm{MB}$ gravity \cite{Blagojevic05} in terms of the charges $\mathcal L_\pm,\mathcal W_\pm$. Since the couplings for teleparallel and $\mathrm{EH}$ gravity coincide, their entropy are still equal. Furthermore, just as in spin-2 case, the sign rule for negative $k_\pm$ is still valid.

\section{Singular point of spin-3 Mielke-Baekler gravity:}\label{sec4}

In the previous section, we built the spin-3 $\mathrm{MB}$ theory from the $\mathrm{CS}$ formulation. This imposes a condition on the parameters:  $\alpha_3\alpha_4-\mathrm{a}^2\neq0$, since the parameters $l,\mathrm{p},\mathrm{q}$ must exist. Now, let us change the perspective and consider as starting point the action \eqref{HS02} with no restriction on the parameters $\mathrm{a},\Lambda,\alpha_3,\alpha_4$. This is similar to the pure $\mathrm{MB}$ gravity theory \eqref{01} where the parameters were free.

The equations of motion \eqref{hs05a}-\eqref{hs05d} are still valid. However, when
\begin{equation}
\frac{\mathrm a}{\alpha_3} = \frac{\alpha_4}{\mathrm a} = - \frac{\Lambda}{\alpha_4} \equiv \mu \label{c01}
\end{equation}
they will reduce to only two:
\begin{eqnarray}
R^{(3)}_{a}+\mu T^{(3)}_{a}+\mu^2 \epsilon_{abc}\left[ \frac{1}{2}e^{b}\wedge e^{c}-2\sigma e^{bd}\wedge e_d^{\;c}\right] &=& 0  \label{c02a} \\
R^{(3)}_{ab}+\mu T^{(3)}_{ab}+\mu^2 \epsilon_{cd(a}e^{c}\wedge e_{b)}^{\; d} &=& 0 \label{c02b}
\end{eqnarray}
meaning that we need further information to determine $R^{(3)}_{a},T^{(3)}_{a}$ and $R^{(3)}_{ab},T^{(3)}_{ab}$. Furthermore, for the singular point \eqref{c01} we have $\alpha_3\alpha_4 - \mathrm{a}^2=0$. Therefore, it is outside the $\mathrm{AdS}$ sector and cannot be written as the difference of two $\mathrm{CS}$ actions.

We can follow a similar procedure as the one performed in \cite{Santamaria} in the context of pure $\mathrm{MB}$ gravity. Since at the singular point there is not enough information about the curvature and torsion components, we can impose vanishing torsion $\mathrm{p}=0$. This can be accomplished from the $\mathrm{CS}$ sector if we write
\begin{equation}
\alpha_{3}\alpha_{4}-\mathrm{a}^{2}	=\varepsilon,\;\;\;
\mathrm{a}\Lambda+\left(\alpha_{4}\right)^{2}=\frac{\varepsilon}{l^{2}},\;\;\ \alpha_{3}\Lambda+\alpha_{4}\mathrm{a}=0\label{c02c}
\end{equation}
where $\varepsilon$ is a parameter, that will eventually go to zero to reach the singular point. Under \eqref{c02c}, the $\mathrm{CS}$ levels become, when $\varepsilon\rightarrow0$: $k_+=8\pi\mathrm{a}l$ and $k_-=0$, meaning that the two $\mathrm{CS}$ levels of the original theory degenerate to one. In addition, we still need to introduce Lagrange's multipliers due to the zero torsion condition, making this model similar to the chiral point of spin-3 topologically massive gravity. In the following section, we will use the canonical formalism to understand the singular point without imposing further information on the torsion.

\subsection{Canonical analysis at the singular point:}

Since the structure of spin-3 $\mathrm{MB}$ gravity is no longer the difference of two $\mathrm{CS}$ actions, we need to analyse its canonical structure. The canonical analysis of pure $\mathrm{MB}$ gravity has been performed in \cite{canonicalMB}, at its singular point in \cite{Santamaria} and the canonical analysis of topologically massive gravity at the chiral point in \cite{GrumillerTMG}. It is also worth noticing that the canonical charges of topologically massive gravity has been computed in \cite{Miskovic2}.

Let us begin the canonical analysis introducing a foliation of space-time: We denote the temporal coordinate by $x^0=t$ and by $x^i$ the angular and radial coordinates. After eliminating some boundary terms, it is possible to write the spin-3 $\mathrm{MB}$ Lagrangian as
\begin{eqnarray}
\mathcal{L}_{\mathrm{MB}}^{\mathrm{Spin-3}}	&=&	 \mathrm{a}\mu\varepsilon^{ij}e_{j}^{a}\partial_{0}e_{ai}-2\sigma\mathrm{a}\mu\varepsilon^{ij}e_{j}^{ab}\partial_{0}e_{abi}+2\mathrm{a}\varepsilon^{ij}\left(e_{j}^{a}+\frac{1}{2\mu}\omega_{j}^{a}\right)\partial_{0}\omega_{ai}\nonumber\\
&&-4\mathrm{a}\varepsilon^{ij}\sigma\left(e_{j}^{ab}+\frac{1}{2\mu}\omega_{j}^{ab}\right)\partial_{0}\omega_{abi}-\left(e_{0}^{a}+\frac{1}{\mu}\omega_{0}^{a}\right)\mathcal{J}_{a}-\left(e_{0}^{ab}+\frac{1}{\mu}\omega_{0}^{ab}\right)\mathcal{J}_{ab} \label{c03a}
\end{eqnarray}
where the quantities $\mathcal{J}_a,\mathcal{J}_{ab}$ are defined by
\begin{eqnarray}
\mathcal{J}_{a} &\equiv& -\mathrm{a}\varepsilon^{ij}\left[R_{aij}+\mu T_{aij}+2\mu^{2}\epsilon_{abc}\left(\frac{1}{2}e_{i}^{b}e_{j}^{c}-2\sigma e_{\; di}^{b}e_{j}^{dc}\right)\right]\label{c04a}\\
\mathcal{J}_{ab} &\equiv& 2\mathrm{a}\sigma\varepsilon^{ij}\left[R_{abij}+\mu T_{abij}+\mu^{2}\epsilon_{mn(a}e_{i}^{m}e_{b)j}^{\; n}-\mu^{2}\epsilon_{mn(a}e_{j}^{m}e_{b)i}^{\; n}\right]\label{c04b}.
\end{eqnarray}
Note that these quantities are related to the left side of the equations of motion \eqref{c02a}, \eqref{c02b}. Then, we must expect, from the Hamiltonian analysis, that $\mathcal{J}_a,\mathcal{J}_{ab}$ are zero.

Let us denote $(\pi^{a\alpha},\pi^{ab\alpha},\Pi^{a\alpha},\Pi^{ab\alpha})$ as the canonical momenta of $(e_{a\alpha},e_{ab\alpha},\omega_{a\alpha},\omega_{ab\alpha})$, respectively. Then, since there are no temporal derivatives of the fields $(e_{a0},e_{ab0},\omega_{a0},\omega_{ab0})$, we have the following primary constraints:
\begin{equation}\label{c05}
\phi^{a0}	\equiv	\pi^{a0}\approx0,\;\;\;\phi^{ab0}	\equiv	\pi^{ab0}\approx0,\;\;\;\Phi^{a0}	\equiv	 \Pi^{a0}\approx 0,\;\;\;
\Phi^{ab0}	\equiv	\Pi^{ab0}\approx0
\end{equation}
and from the definition of canonical momenta of the remaining fields we have
 \begin{eqnarray}
\phi^{ai}	\equiv	\pi^{ai}-\mathrm{a}\mu\varepsilon^{ij}e_{j}^{a}\approx 0, \;\;\;&& \Phi^{ai}	\equiv	 \Pi^{ai}-2\mathrm{a}\varepsilon^{ij}\left(e_{j}^{a}+\frac{1}{2\mu}\omega_{j}^{a}\right)\approx 0\label{c06a}\\
\phi^{abi}	\equiv	\pi^{abi}+2\mathrm{a}\sigma\mu\varepsilon^{ij}e_{j}^{ab}\approx 0,\;\;\;&&
\Phi^{abi}	\equiv	 \Pi^{abi}+4\mathrm{a}\sigma\varepsilon^{ij}\left(e_{j}^{ab}+\frac{\mathrm{1}}{2\mu}\omega_{j}^{ab}\right)\approx 0.\label{c06b}
\end{eqnarray}
As we see, all canonical momenta will define primary constraints, which simplifies the canonical analysis. The primary Hamiltonian is given by:
\begin{equation}\label{c07}
\mathcal H_T=\left(e_{0}^{a}+\frac{1}{\mu}\omega_{0}^{a}\right)\mathcal{J}_{a}+\left(e_{0}^{ab}+\frac{1}{\mu}\omega_{0}^{ab}\right)\mathcal{J}_{ab}+u_{a\alpha}\phi^{a\alpha}+u_{ab\alpha}\phi^{ab\alpha}+v_{a\alpha}\Phi^{a\alpha}+v_{ab\alpha}\Phi^{ab\alpha}
\end{equation}
where the $u$'s and $v$'s are Lagrange multipliers.
The consistency condition of the four constraints \eqref{c05} lead us to only two secondary constraints
\begin{equation}\label{c08}
    \mathcal{J}_{a}\approx 0, \;\;\; \mathcal{J}_{ab}\approx 0
\end{equation}
as predicted, and in perfect agreement with the equations of motion. The consistency condition for \eqref{c06a}, \eqref{c06b} gives, again, only two relations:
\begin{eqnarray}
\partial_{0}\omega_{i}^{a}+\mu\partial_{0}e_{i}^{a}-\mu u_{i}^{a}-v_{i}^{a} &\approx& 0\label{c09a}\\
\partial_{0}\omega_{i}^{ab}+\mu\partial_{0}e_{i}^{ab}-\mu u_{i}^{ab}-v_{i}^{ab}&\approx&0\label{c09b}.
\end{eqnarray}
where we used the equations of motions to simplify the expressions. Note that we can solve the above equations for $(v^a_i,v^{ab}_i)$ or for $(u^a_i,u^{ab}_i)$. This indeterminacy is related to the singular matrix build with the Poisson brackets of the constraints \eqref{c06a}, \eqref{c06b}. The next step in the Dirac's algorithm is to test the consistency conditions for the secondary constraints \eqref{c08}. Since the Poisson Brackets between $(\mathcal J_a,\mathcal J_{ab})$ with the constraints \eqref{c06a}, \eqref{c06b} are non-zero, it looks like there will be new relations between the parameters. However, when we replace \eqref{c09a}, \eqref{c09b}, we notice that the consistency conditions are satisfied. Therefore, the algorithm is closed.

We can now replace \eqref{c09a}, \eqref{c09b} in the Primary Hamiltonian. After eliminating some boundary terms we obtain
\begin{eqnarray}
\mathcal H_T &=&  \left(e_{0}^{a}+\frac{1}{\mu}\omega_{0}^{a}\right)\bar{\mathcal{J}_{a}}+\left(e_{0}^{ab}+\frac{1}{\mu}\omega_{0}^{ab}\right)\bar{\mathcal{J}_{ab}}+u_{ai}\chi^{ai}+u_{abi}\chi^{abi}\nonumber \\
&& + u_{a0}\phi^{a0}+u_{ab0}\phi^{ab0}+v_{a0}\Phi^{a0}+v_{ab0}\Phi^{ab0}\label{c10}
\end{eqnarray}
where
\begin{eqnarray}
\bar{\mathcal{J}_{a}} &\equiv& \mathcal{J}_{a}-\mu\partial_{i}\Phi_{a}^{i}-\mu\epsilon_{abc}\left(\omega_{i}^{b}+\mu e_{i}^{b}\right)\Phi^{ci}-2\mu\epsilon_{abc}\left(\omega_{qi}^{b}+\mu e_{qi}^{b}\right)\Phi^{qci}\label{c11a}\\
\bar{\mathcal{J}_{ab}} &\equiv& \mathcal{J}_{ab}-\mu\partial_{i}\Phi_{ab}^{i}-\mu\epsilon_{mn(a}\left(\omega_{i}^{m}+\mu e_{i}^{m}\right)\Phi_{b)}^{ni}+2\sigma\mu\epsilon_{mn(a}\left(\omega_{b)i}^{\; m}+\mu e_{b)i}^{\; m}\right)\Phi^{ni}\label{c11b}
\end{eqnarray}
and
\begin{equation}
\chi^{ai} \equiv \phi^{ai}-\mu\Phi^{ai}\;\;\;\;\;\chi^{abi} \equiv \phi^{abi}-\mu\Phi^{abi}.\label{c12}
\end{equation}
Note that the constraints \eqref{c12} appear naturally after replacing the Lagrange's multipliers $v^a_i,v^{ab}_i$.

Now, it is time to classify the constraints. The set $\left(\Phi^{ai},\Phi^{abi}\right)$ is second-class, since
\begin{equation}\label{c13}
\left\{ \Phi^{ai}(x),\Phi^{bj}(y)\right\} =-\frac{2\mathrm{a}}{\mu}\varepsilon^{ij}\eta^{ab}, \;\;\;\;\left\{ \Phi^{abi}(x),\Phi^{cdj}(y)\right\} = \frac{2\mathrm{a}\sigma}{\mu}\varepsilon^{ij}\eta^{a(c}\eta^{d)b}
\end{equation}
The set $\left(\phi^{a0},\phi^{ab0},\Phi^{a0},\Phi^{ab0},\chi^{ai},\chi^{abi},\bar{\mathcal J}^a,\bar{\mathcal J}^{ab}\right)$ is first-class, since the non-zero Poisson brackets close an algebra
\begin{eqnarray}
\left\{ \bar{\mathcal J}_a(x),\bar{\mathcal J}_b(y)\right\} &=& -\mu \epsilon_{ab}^{\;\;\; c}\bar{\mathcal J}_c(x)\label{c14a}\\
\left\{ \bar{\mathcal J}_a(x),\bar{\mathcal J}_{bc}(y)\right\} &=& -\mu\epsilon_{d}^{\; a(b}\mathcal{\bar{J}}^{c)d}\left(x\right)\label{c14b}\\
\left\{ \bar{\mathcal J}_{ab}(x),\bar{\mathcal J}_{cd}(y)\right\} &=& -\mu\sigma\left(\eta^{a(c}\epsilon_{\;\;\;\; m}^{d)b}+\eta^{b(c}\epsilon_{\;\;\;\; m}^{d)a}\right)\bar{\mathcal{J}}^{m}\left(x\right).\label{c14c}
\end{eqnarray}
Finally, the Poisson Brackets between the two sets are
\begin{eqnarray}
\left\{ \bar{\mathcal{J}}^{a}\left(x\right),\Phi^{bi}\left(y\right)\right\} 	=	-\mu\epsilon_{\;\;\; c}^{ab}\Phi^{ci}\left(x\right), &\;\;\;&
\left\{ \bar{\mathcal{J}}^{a}\left(x\right),\Phi^{bci}\left(y\right)\right\} 	=	-\mu\epsilon_{d}^{\; a(b}\Phi^{c)di}\left(x\right)\label{c15a}\\
\left\{ \bar{\mathcal{J}}^{ab}\left(x\right),\Phi^{ci}\left(y\right)\right\} = \mu\epsilon_{d}^{\; c(a}\Phi^{b)di}\left(x\right),  &\;\;\;& \left\{ \bar{\mathcal{J}}^{ab}\left(x\right),\Phi^{cdi}\left(y\right)\right\}=\sigma\mu\epsilon_{mn}^{\;\;\; (a}\eta^{b)(c}\eta^{d)m}\Phi^{ni}\left(x\right)\label{c15b}
\end{eqnarray}
where the Dirac's delta has been omitted in the above brackets. The second-class constraints can be eliminated by building Dirac's brackets. This way $\Phi^{ai}=0$ and $\Phi^{abi}=0$ can be used as strong equations. We will not perform that task since the important result lies in the algebra of the first-class constraints \eqref{c14a}-\eqref{c14c} which, after an appropriate scaling, becomes a single copy of the $\mathfrak{sl}(3)$ algebra \eqref{HS01a}-\eqref{HS01c}. This was expected from the limit \eqref{c02c}. However, with the canonical analysis we did not impose any further condition on the torsion components, as shown explicitly by the constraints \eqref{c04a}-\eqref{c04b}.

Finally, notice that the dimensions of the phase-space is $N=108$, which is the counting of the variables of the spin-2 sector $\left(e_{a\mu},\omega_{a\mu}\right)$, spin-3 sector $\left(e_{ab\mu},\omega_{ab\mu}\right)$, and their respective canonical momenta. We have $S=18$ second-class constraints, and $M=45$ first-class constraints. This gives a total of $N-2M-S=0$ degrees of freedom, as expected from a topological theory.

\section{Final considerations} \label{sec5}

In this work we presented the spin-3 Mielke-Baekler gravity theory \eqref{HS02} with four couplings, which allow us to build the higher-spin versions of the Einstein-Hilbert, Teleparallel and Exotic gravity. We performed this generalization with the aid of the Chern-Simons formulation of Mielke-Baekler pure gravity, following a similar approach to the spin-3 generalizations of Einstein-Hilbert gravity and Topologically Massive gravity. We showed that, in the presence of non-vanishing torsion components $T^{(3)}_a$, $T^{(3)}_{ab}$, the asymptotic symmetries are two copies of the $W_3$ algebra with different central charges \eqref{BH06}. It is worth noticing that boundary condition \eqref{BH03} satisfies the so-called highest weight ansatz. However, as shown in \cite{Grumiller3D} in the context of $\mathrm{EH}$ gravity (and also in flat gravity \cite{Flat}, higher-spin \cite{Chethan}, supergravity \cite{Valcarcel} and higher-spin supergravity \cite{Ozer}), it is possible to consider a different ansatz that will lead to new asymptotic symmetries, such as a higher-spin generalization of \cite{NHTorsion}.

Since the pure Mielke-Gravity theory is not restricted to its $\mathrm{AdS}$ sector, we eliminated the previous conditions on the couplings for the spin-3 theory. Then, outside the $\mathrm{AdS}$ sector we found a singular point where the number of equations of motions is reduced in half and we cannot determine neither torsion or curvature components. This degeneracy is also present in the canonical analysis, where some consistency conditions are redundant, since they reproduce the same constraints. The final counting of degrees of freedom is zero, meaning that the theory at the singular point is still topological.

As a future perspective, we will study the thermodynamic properties of other spin-3 $\mathrm{MB}$ black-hole solution \cite{Castro} and analyse how torsion modifies the black-hole entropy. In particular, it is known that Exotic gravity has reversed roles for mass and angular momentum, it will be interesting to see if this characteristic is still preserved in the presence of spin-3 fields.

\section{Acknowledgements}
The authors thank M. C. Bertin for reading the manuscript and suggestions. J. R. B. Peleteiro thanks Capes for financial support. C. E. Valc\'arcel thanks J. Chopin for discussions at \textbf{Encontro Geom\'etrico} meetings.

\section{Appendix}
We are using the following representation for the $\mathfrak{sl}(2)$ algebra
\begin{equation}
L_{-1}=\left(\begin{array}{cc}
0 & -1\\
0 & 0
\end{array}\right),\;\; L_{0}=\left(\begin{array}{cc}
\frac{1}{2} & 0\\
0 & -\frac{1}{2}
\end{array}\right),\;\; L_{1}=\left(\begin{array}{cc}
0 & 0\\
1 & 0
\end{array}\right),
\end{equation}
and for the $\mathfrak{sl}(3)$ algebra we follow the matrix representation of \cite{Campoleoni}:
\begin{equation}
L_{1}=\left(\begin{array}{ccc}
0 & 0 & 0\\
1 & 0 & 0\\
0 & 1 & 0
\end{array}\right),\;\;\;
L_{0}=\left(\begin{array}{ccc}
1 & 0 & 0\\
0 & 0 & 0\\
0 & 0 & -1
\end{array}\right),\;\;\;
L_{-1}=\left(\begin{array}{ccc}
0 & -2 & 0\\
0 & 0 & -2\\
0 & 0 & 0
\end{array}\right)
\end{equation}
and
\begin{eqnarray}
W_{2}=2\sqrt{-\sigma}\left(\begin{array}{ccc}
0 & 0 & 0\\
0 & 0 & 0\\
1 & 0 & 0
\end{array}\right),\;\;\;
W_{1}=\sqrt{-\sigma}\left(\begin{array}{ccc}
0 & 0 & 0\\
1 & 0 & 0\\
0 & -1 & 0
\end{array}\right),\;\;\;
W_{0}=\frac{2}{3}\sqrt{-\sigma}\left(\begin{array}{ccc}
1 & 0 & 0\\
0 & -2 & 0\\
0 & 0 & 1
\end{array}\right)\\
W_{-1}=\sqrt{-\sigma}\left(\begin{array}{ccc}
0 & -2 & 0\\
0 & 0 & 2\\
0 & 0 & 0
\end{array}\right),\;\;\;
W_{-2}=2\sqrt{-\sigma}\left(\begin{array}{ccc}
0 & 0 & 4\\
0 & 0 & 0\\
0 & 0 & 0
\end{array}\right).
\end{eqnarray}

\end{document}